\documentclass[letter]{aa} 

\usepackage{graphicx}
\usepackage{txfonts}
\begin{document} 

   \title{A too-many dwarf galaxy satellites problem in the M\,83 group }
\author{Oliver M\"uller\inst{1,2} \and Marcel S. Pawlowski\inst{3} \and  Yves Revaz\inst{1} \and  Aku Venhola\inst{4} \and  Marina Rejkuba\inst{5} \and  Michael Hilker\inst{5} \and  Katharina Lutz\inst{2}}

 \institute{Institute of Physics, Laboratory of Astrophysics, Ecole Polytechnique F\'ed\'erale de Lausanne (EPFL), 1290 Sauverny, Switzerland\\
  \email{oliver.muller@epfl.ch}
 \and
 Universit\'e de Strasbourg, CNRS, Observatoire astronomique de Strasbourg (ObAS), UMR 7550, 67000 Strasbourg, France
 \and 
 Leibniz-Institut fur Astrophysik Potsdam (AIP), An der Sternwarte 16, D-14482 Potsdam, Germany
\and
Space physics and Astronomy Research Unit, University of Oulu, Oulu, FI-90014 Finland
  \and 
 European Southern Observatory, Karl-Schwarzschild Strasse 2, 85748, Garching, Germany
}
   \date{Received tba; accepted tba}

  \abstract{Dwarf galaxies in groups of galaxies provide excellent test cases for models of structure formation. This led to a so-called small-scale crisis, including the famous {missing satellite} and {too-big-to-fail} problems. It was suggested that these two problems are solved by the introduction of baryonic physics in cosmological simulations. We test  for the nearby grand spiral M\,83 -- a Milky Way sibling -- whether its number of dwarf galaxy companions is compatible with today's $\Lambda$ + Cold Dark Matter model using two methods: with cosmological simulations that include baryons, as well as with theoretical predictions from the sub-halo mass function. 
By employing distance measurements we recover a list of confirmed dwarf galaxies within 330\,kpc around M\,83 down to a magnitude of $M_V =-10$. We found that both the state-of-the-art hydrodynamical cosmological simulation Illustris-TNG50 and theoretical predictions {agree with the number of confirmed satellites around M\,83 at the bright end of the luminosity function ($>$10$^8$ solar masses) but underestimate it at the faint end (down to 10$^6$ solar masses) at more than 3$\sigma$ and 5$\sigma$ levels, respectively}. This indicates a {too-many satellites} problem in $\Lambda$CDM for M\,83. The actual degree of tension to cosmological models is underestimated, because the number of observed satellites is incomplete due to the high contamination of spurious stars and galactic cirrus.} 

   \maketitle
%

\section{Introduction}

In recent few years,  the 
task of improving and extending the census of dwarf galaxies in the nearby universe has received an immense boost. This is thanks to the development and improvement of instruments and facilities \citep[e.g., ][]{2003AJ....126.2081A,
2004MNRAS.350.1195M,2008PASP..120..212G,2010SPIE.7735E..08B,2014PASP..126...55A,2018PASJ...70S...1M,2019AJ....157..168D}, as well as the general interest in these objects in the context of cosmology and galaxy formation \citep[e.g., ][]{2010A&A...523A..32K,2012A&A...541A.131C,2015MNRAS.446..120D,2015PNAS..11212249W,2016PDU....12...56B,Bullock2017,2018Natur.563...85H,2018Natur.555..629V,2021MNRAS.502.5921S,2021NatAs...5.1185P,2022NatAs...6..897S,2023MNRAS.524..952K,2023A&A...673A.160M}. Dwarf galaxy 
abundance gives valuable insights into models of structure formation. The apparent discrepancies between the predicted number of subhalos and the observed Milky Way satellites by dark-matter-only cosmological simulations, i.e. the missing-satellite problem \citep{1999ApJ...524L..19M,2008ApJ...688..277T} and too-big-to-fail problem \citep{2011MNRAS.415L..40B,2014MNRAS.444..222G}, has motivated to improve our understanding of the role of baryons within galaxies \citep{2007ApJ...670..313S,2016ApJ...827L..23W}. Today, the observed Luminosity Function (LF) of the Milky Way {and Andromeda satellites} can be reproduced by cosmological simulations, when baryonic effects like supernova feedback are  included \citep{2016MNRAS.457.1931S,2018MNRAS.478..548S,2018A&A...616A..96R}. However, the question remains whether these results extend and remain valid outside of the Local Group, especially because the Local Group served to calibrate the baryonic effects, making these successes not independent from the observations. 
Other problems, like a tension in the motion and distribution of the satellites around the Milky Way and Andromeda, are still open to debate \citep{2013Natur.493...62I,2015MNRAS.452.1052L,2018MPLA...3330004P,2021NatAs...5.1185P,2023NatAs...7..481S}.

To see how typical our Local Group is, dwarf galaxies in other nearby groups need to be discovered and their memberships to the host galaxies have to be accurately established. 
This is a major undertaking. The membership determination requires identification of small low surface brightness galaxies and accurate distance measurements for these \citep{2022ApJ...933...47C}. However, 
getting this distance information is observationally expensive at the faint end of the LF, especially when there is a large number of dwarf galaxy candidates to follow up \citep[e.g., ][]{2009AJ....137.3009C,2016A&A...588A..89J,2018A&A...615A.105M,2020MNRAS.491.1901H,2023MNRAS.tmp.3597C}. Using the tip of the red giant branch (TRGB, \citealt{daCostaArmandroff1990,Lee1993}) as a standard candle it takes one orbit  per target for galaxies within $<10$\,Mpc with the Hubble Space Telescope (HST, e.g., \citealt{2007AJ....133..504K,2013AJ....146..126C,2019ApJ...872...80C}), or several hours of observation with ground based 8-meter class telescopes, such as the Very Large Telescope (VLT, see e.g., \citealt{2015A&A...575A..72B}). 
Determining distances {via period-luminosity relations \citep{1912HarCi.173....1L} of} variable stars that can be used as distance indicators is {more difficult} due to the {need for multi-epoch observations combined with few (if any)} young Cepheid variables and faint magnitudes of RR Lyrae {in dwarf galaxies}. In particular the latter are only observable out to distances of about 2\,Mpc with the HST \citep{2010ApJ...708L.121D}. Type II Cepheids are brighter and could thus be observed to slightly larger distances in dwarf galaxies hosting intermediate age and old stellar populations. However, they are less numerous than RR Lyrae, and have so far only been studied within the Local Group \citep{2020JApA...41...23B}. \citet{2023arXiv231201420B} recently suggested that the velocity dispersion of globular clusters is an excellent distance indicator on par with the TRGB. They independently  measured the distance of Centaurus A (Cen\,A) which is consistent with previous distance estimates from other distance indicators. However, getting high-resolution spectroscopy for globular cluster systems is an expensive undertaking and may not always be feasible.

Another {way} to measure distances -- which is currently only accurately applicable for early-type galaxies -- is the surface brightness fluctuation (SBF) method \citep{1988AJ.....96..807T}. While not as accurate as the TRGB, it can be used on sufficiently deep enough images to {estimate}
the distance and with that, the membership of the dwarf galaxies to a potential host. With this technique, several of the dwarf galaxy candidates found around M\,101 \citep{2016A&A...588A..89J,2017A&A...602A.119M,2017ApJ...850..109B} could either be confirmed or excluded \citep{2019ApJ...878L..16C}. \citet{2022ApJ...933...47C} successfully extended this SBF analysis exploiting public data to other giant galaxies in the nearby universe, showing that the SBF method is a serious alternative to confirm/reject membership of dwarf galaxies in group environments. However, we also note that when it comes to precise distance measurements, SBF may fail (see e.g. \citealt{2001A&A...371..487J} for a direct comparison of the SBF and TRGB distance measurement of the same dwarf galaxy).

In terms of cosmological predictions the abundance of dwarf galaxies in 
nearby groups  has been addressed in a few recent studies.
Around M\,94 -- a giant spiral galaxy at the heart of the Canes Venatici I cloud -- an apparent lack of dwarf galaxy satellites was found \citep{2018ApJ...863..152S}. Employing a deep imaging campaign with the Hyper Suprime Cam and resolving the upper part of the red giant branch \citet{2018ApJ...863..152S} surveyed the 150\,kpc vicinity of M\,94 and found only two satellites. Their membership was established with the TRGB. This is in stark contrast to the expected number of satellites  of five to ten, begging the question of whether the missing-satellite problem is back again.
On the other hand, for Cen\,A -- a giant elliptical galaxy,  which has been the subject of several dedicated deep surveys \citep{2014ApJ...795L..35C,2016ApJ...823...19C,2019ApJ...872...80C,2017A&A...597A...7M,2019A&A...629A..18M,2021A&A...645A..92M,2017MNRAS.469.3444T,2018ApJ...867L..15T} -- the LF within 200\,kpc of Cen\,A is well compatible within 2$\sigma$ with cosmological simulations including baryonic physics \citep{2019A&A...629A..18M}. Similarly, for low-mass host galaxies in the local galactic neighborhood, as well as the high redshift universe,  the LF agrees well with cosmological predictions, when biases in observations are taken into account \citep{2020A&A...644A..91M,2021MNRAS.502.1205R}. \citet{2021ApJ...908..109C} used the stellar-halo mass relation to predict the number of satellites for 12 giant galaxies in the nearby universe (incl. Cen\,A and M\,94) and compared it to observations. They found a large scatter in the luminosity function, which is in agreement with cosmology, but pointed  out that on the bright and faint ends there are more observed dwarfs than expected. \citet{2023MNRAS.521.4009C} investigated the luminosity function of NGC\,2683 and other nearby giant galaxies with respect to the brightest dwarfs. Studying the magnitude gap between the host galaxy and the brightest satellite, they found that three out of six systems have a larger magnitude gap than expected in cosmological simulations (namely TNG100 of the Illustris suit of simulations, \citealt{2018MNRAS.475..676S}), meaning that the brightest satellites may be missing. 

These ambiguous results show the urgent need for complete dwarf galaxy samples within nearby galaxy groups to study the abundance of satellites. Here, we present a census of the spiral galaxy M\,83 (NGC\,5236) -- also known as the Southern Pinwheel Galaxy -- which is one of the closest neighbours to our Local Group. This late-type galaxy is at a distance of $D\simeq4.9$ Mpc \citep{2008ApJ...683..630H}, and with Cen\,A at $D\simeq3.8$ Mpc \citep{2004A&A...413..903R} it forms the Centaurus group, similar to the Local Group. 
Several of Cen\,A's satellite members are in projection closer to M\,83, thus there is a potential of confusion which requires distance measurements to resolve. The mass of M\,83 has been studied by different groups. Recently, the flat part of the rotation curve (tracing its mass) was updated from with  $v_{flat}\approx 150$\,km\,s$^{-1}$ \citep{2015MNRAS.452.3139K} to $v_{flat}\approx 190$\,km\,s$^{-1}$ \citep{2021A&C....3400448D}. This is compatible with the rotation curve of the Milky Way, indicating that the two systems have similar masses and should therefore follow the same scaling relations when it comes to the number of expected dwarf galaxy satellites \citep{2019ApJ...870...50J}.

\section{Luminosity function of the M\,83 group}

\begin{figure*}
\centering
\includegraphics[width=\linewidth]{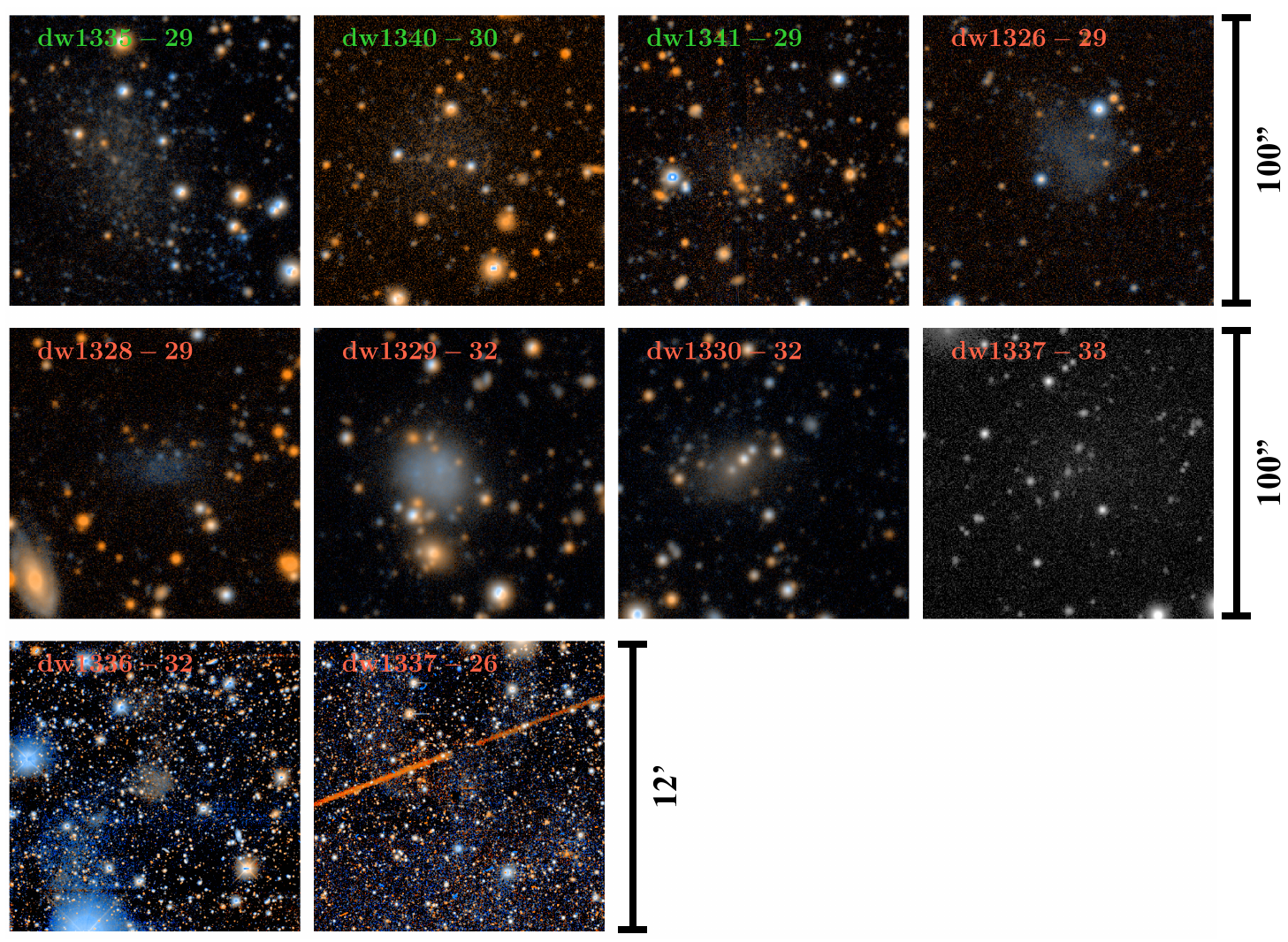}
 \caption{\label{fig:sbf} Color-composite images of previous dwarf galaxy candidates of M\,83 {which have an optical counterpart in deeper images}. With green labels the galaxies featuring an SBF are shown, in red objects without any SBF signal.}
\end{figure*}

\begin{figure}
\centering
\includegraphics[width=\linewidth]{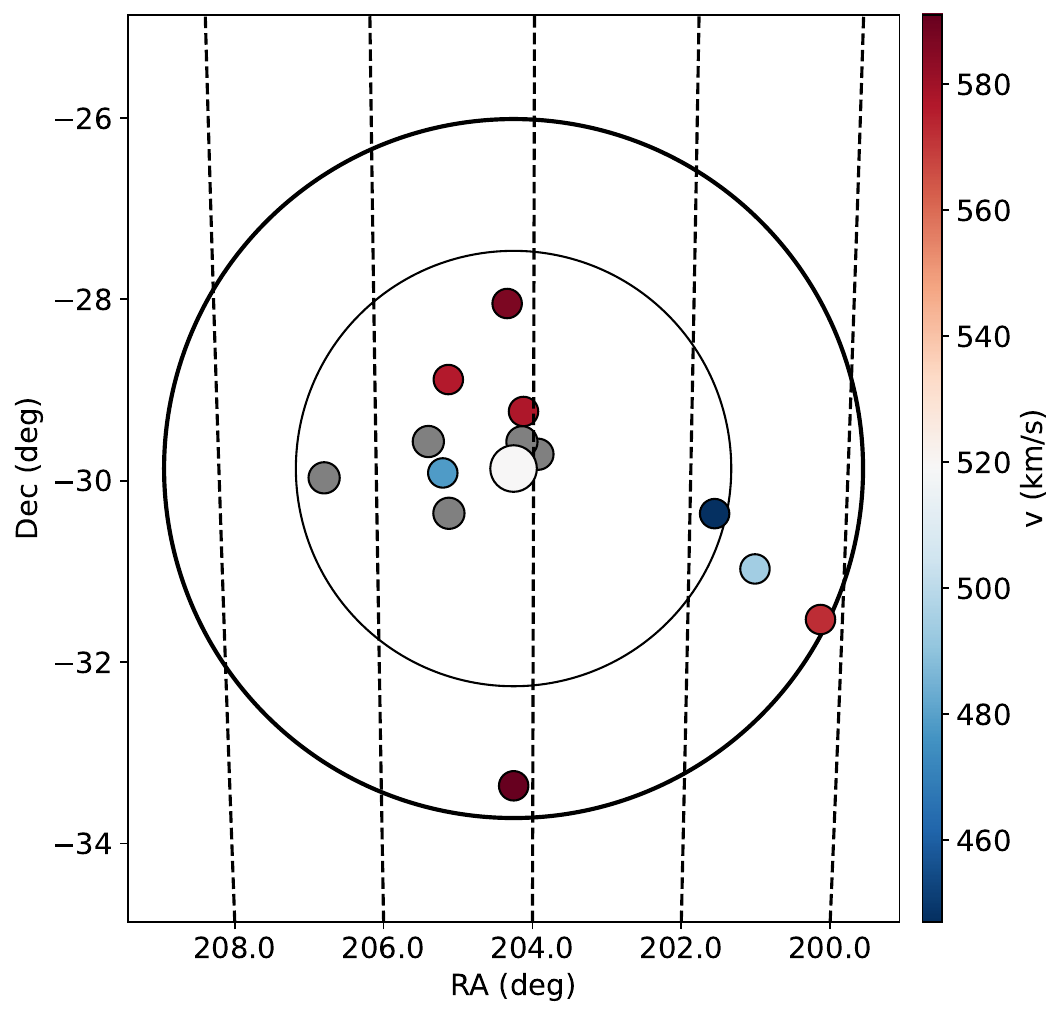}
 \caption{\label{fig:m83}The field around M\,83. The large dot corresponds to M\,83, the small dots to confirmed dwarf galaxies. In color are dwarf galaxies with velocity measurements, in gray without. The thin circle corresponds to the virial radius of M\,83, the large circle to 330\,kpc.}
\end{figure}

\begin{table*}[ht]
\centering                          
\caption{{Dwarf galaxies within a projected radius of 330\,kpc around M\,83.}}
\begin{tabular}{l c c c r c c}        
\hline\hline                 
Name & $\alpha$ & $\delta$ &   $D$ &$\Delta_{2D}$ &  v & $M_V$\\    
 & J2000.0 & J2000.0  & Mpc & kpc & km/s & mag \\  
\hline      \\[-2mm]                  
KK\,195 & 13:21:08.2  & 	$-$31:31:47 & 5.22 & 325 &  572 &$-$12.1 \\
KK\,200 &	13:24:36.0	&$-$30:58:20	&	4.76 & 248 & 494 &	$-$13.5\\
IC\,4247 &	13:26:44.4 	& $-$30:21:45 & 5.18 & 195 & 420 & $-$15.0 \\
dw1335-29 & 13:35:46.7 & $-$29:42:28 & 5.03 & 27 &  --- & $-$10.3 \\
UGCA\,365 & 13:36:30.8 & $-$29:14:11 & 5.42 & 55 &  577 &$-$14.1 \\
KK\,208 & 13:36:35.5 	 & $-$29:34:15 & 5.01 & 26 &  --- & ($-$15.9) \\
HIDEEP J1337-33 & 13:37:00.6 &	$-$33:21:47 & 4.55 & 299 &  591 &$-$11.3 \\
ESO\,444-084 & 13 37 20.2 &	$-$28:02:46 & 4.61 & 156 &  587 &$-$14.1 \\
 dw1340-30 & 13:40:19.2	 & $-$30:21:31 & 5.06 & 74 &  --- &$-$10.8 \\
IC\,4316 & 13:40:18.1 	 & $-$28:53:40 & 4.35 & 104 &  576 &$-$15.2 \\
dw1341-29 & 13:41:20.2 & $-$29:34:03 & $\approx$4.9 & 84 &  --- & $-$8.8 \\
NGC5264& 13:41:37.0 & $-$29:54:50 & 4.79 & 67 &  478 & $-$16.7 \\
KK218/Cen\,A-dE4 & 13:46:39.5  & 	$-$29:58:45 & 4.94 &  178&  --- &$-$12.1 \\
\hline 
\end{tabular}
\tablefoot{The TRGB distances and velocities are compiled in the LV catalog \citep{2013AJ....145..101K} and are from the HST programs of \citep{2002A&A...385...21K,2007AJ....133..504K,2003ApJ...596L..47P}; and \citep{2007MNRAS.374..107G}, and the VLT program of \citep{2018A&A...615A..96M}. Velocities are from \citep{1999ApJ...524..612B,2004AJ....128...16K,2008MNRAS.386.1667B}; and \citep{2008AstL...34..832K}. The photometry comes from \citet{2015A&A...583A..79M}, except for KK208, where we derive the luminosity from the stellar mass estimation of \citet{2014ApJ...789..126B}.}
\label{table:lf} 

\end{table*}

The M83 group has 13 confirmed dwarf galaxies and eight unconfirmed dwarf galaxy candidates  within a projected separation of 330\,kpc \citep{2015A&A...583A..79M}. The most distant confirmed M\,83 satellite -- KK\,195 -- is at a projected distance of $\approx$330\,kpc. Several dwarf galaxy candidates \citep{2015A&A...583A..79M} are within this radius. These candidates are: dw1326-29, dw1329-32, dw1330-32, dw1334-32,  dw1335-33, dw1336-32,  dw1337-33, and dw1337-26 \citep{2015A&A...583A..79M}, and an additional dwarf galaxy candidate -- dw1341-29 -- we serendipitously discovered while exploring the NOAO Science archive. {To get its magnitude, we fitted an exponential profile with Galfit \citep{2002AJ....124..266P}.}
For all the unconfirmed dwarf galaxy candidates, we apply the SBF method to measure their distance and confirm or reject their memberships to M\,83 in the next paragraph.

While in the NOAO Science archive there is not a deep homogeneous data set covering the whole 330\,kpc, there are several pointings to test whether we can find a SBF distance for the dwarf candidates. In the following we discuss each candidate with respect to its potential SBF signal.
The dwarf candidates dw1334-32 and dw1335-33 are too faint and to low-surface brightness to be able to try measuring SBF, therefore we remove them from our confirmed dwarf galaxy list. {However, they were already listed as unlikely candidates in \citep{2015A&A...583A..79M} and are likely false positives -- both are not visible in deeper images from the The Dark Energy Camera Legacy Survey (DECaLS, \citealt{2019AJ....157..168D}) and
 dw1335-33 has a suspicious edge in the luminosity distribution in Fig.\,2 of \citet{2015A&A...583A..79M} and may be an artefact coming from the stacking of the individual exposures.}
Each of the remaining candidates has combined exposure times ranging from 1310 to 28285\,sec in $g$, $r$, or $i$. These stacked exposures are deep enough to test whether there is a SBF signal consistent with a membership of the M\,83 group. To measure the SBF, we follow the steps as suggested by \citet{2019ApJ...879...13C}.
Only the  dwarf galaxy candidate dw1341-29 shows a signal of being resolved. Its morphology and fluctuation signal resemble the two dwarf spheroidals dw1335-29 and dw1340-30 (see Fig.\,\ref{fig:sbf}) which have been previously confirmed with HST and VLT observations \citep{2017MNRAS.465.5026C,2018A&A...615A..96M}. This dwarf galaxy candidate must therefore be a nearby object.
The dwarf galaxy candidate dw1337-26 looks to be part of a ring of scattered light, therefore it is an artefact.  The dwarf candidate dw1336-32 is part of a cirrus patch. The central part of this cirrus appears as a distinct, round feature, resembling the morphology of a dwarf galaxy. However, faint, adjacent cirrus patches are clearly visible stretching out towards the north and south-east{, see Fig.\ref{cirrus} in Appendix \ref{app:dw1336-32}.}
For the rest, we can confirm that there is no SBF signal visible, meaning that they are real objects but background galaxies that do not belong to Cen\,A or M\,83. This is consistent with recent HI observations covering the region around M\,83, showing that dw1328-29 -- a previously listed dwarf galaxy candidate of M\,83 \citep{2015A&A...583A..79M} --  is a background dwarf galaxy \citep{2019MNRAS.489.5723F}. 
All of these candidates, together with the two  previously confirmed dwarf spheroidals are shown in Fig.\,\ref{fig:sbf}.

The previous assessment of a potential SBF signal leaves us with a census of 13 confirmed dwarf galaxies within 330\,kpc of M\,83 down to at least $-$10\,mag. In Table\,\ref{table:lf} we compile all confirmed dwarf galaxies of M\,83, in Fig.\,\ref{fig:m83} we show their on-sky distribution. {We further show their line-of-sight velocities, which are all within $\pm$100km/s of M\,83's velocity, which can serve as another means of establishing memberships of the group.}
Special attention is required for the tidally disrupted dwarf KK\,208. This object is a stream with an estimated baryonic mass of 1.0 $\times$ 10$^8$ M$_\odot$ \citep{2014ApJ...789..126B} corresponding to a $V$ band magnitude of $\approx$-15.9. This puts it among the brightest satellites in M83. When comparing the M83 satellite system with simulations we test the results by including and excluding this system. We note that the following results do not change if we include or exclude KK\,208.

\section{Comparison to $\Lambda$CDM models}

Now we compare the observed luminosity function of the M\,83 to the predicted number of satellites based on $\Lambda$CDM galaxy formation models using two different methods, this is, a) using a state-of-the art hydrodynamical cosmological simulation, and b) using theoretical predictions from the subhalo-mass function.

\subsection{Illustris-TNG50}

\begin{figure}
\centering
\includegraphics[width=\linewidth]{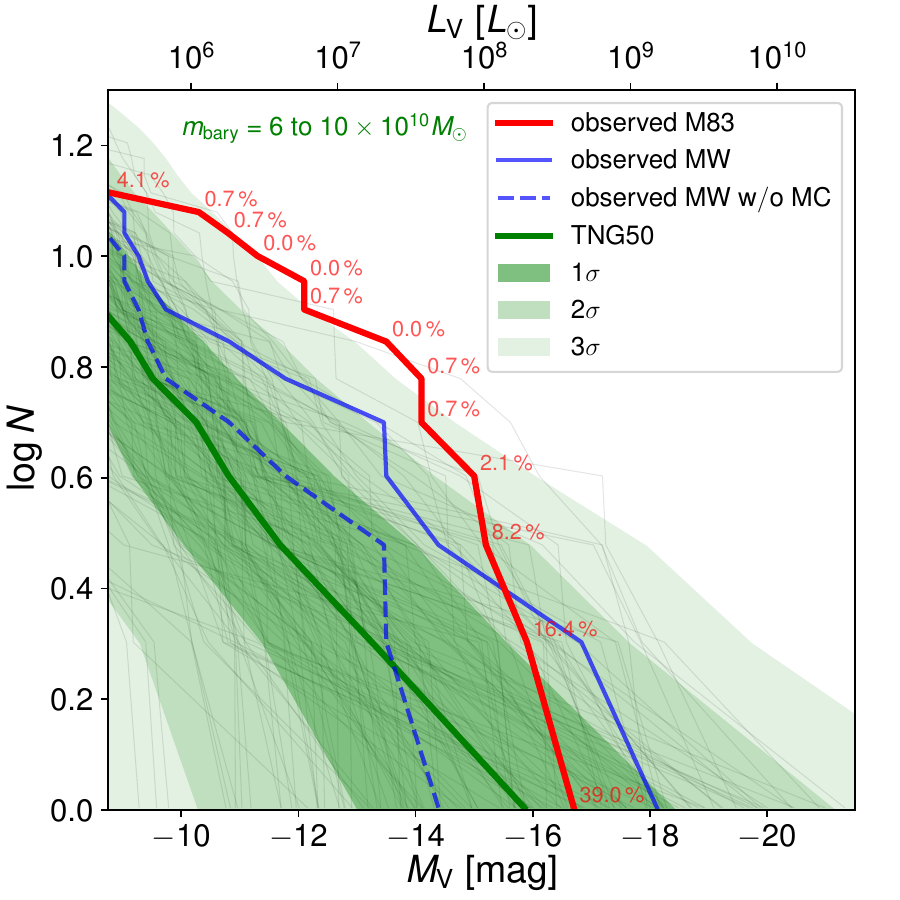}
 \caption{\label{fig:illustris}The observed luminosity function of M\,83 (red line)  compared to its Illustris-TNG50 analogs, with each analog corresponding to a gray line. The 1, 2, and 3$\sigma$ confidence intervals are indicated with the green areas. The red numbers indicate the percentage of analogs having more satellites than the observed one at given step.  Also shown is the Milky Way luminosity function (blue) including (straight) and excluding (dashed) the rare Magellanic Clouds (MC). While the Milky Way luminosity function without the MC  is consistent with expectations from Illustris-TNG50, the M\,83 luminosity function deviates by more than 3$\sigma$. }
\end{figure}

First we use the publicly available $z=0$ galaxy catalogs \citep{2019ComAC...6....2N} from the TNG50-1 simulation of the Illustris-TNG project, a large-scale cosmological simulation including hydrodynamics. 
This run has a box size of 51.7 Mpc and a dark matter particle mass $m_\mathrm{DM} = 4.5 \times 10^5$\,solar masses. 
For the comparison, we follow the same approach as for Cen\,A \citep{2019A&A...629A..18M}. Motivated by the flat part of the rotation curve of M\,83 the baryonic Tully-Fisher relation \citep{2005ApJ...632..859M}, yields a baryonic mass of $0.7\times10^{11}$ solar masses. We select the halo primaries within a baryonic mass range of $0.6\times10^{11} \leq M \leq 1.0\times10^{11}$, which is slightly skewed towards the higher masses, which will result in a larger number of identified subhalos in the simulation but allows us to get a statistically more significant sample. This mass range is also consistent with the estimation of the baryonic content estimated from the $K$-band luminosity ($0.43\times10^{11}$ solar masses \citealt{2013AJ....145..101K}) and HI content ($0.24\times10^{11}$ solar masses \citealt{1981A&A...100...72H}) and includes the baryonic mass estimation of the Milky Way \citep{2020MNRAS.494.4291C}.
We exclude all hosts having another galaxy of the same or higher mass within 700\,kpc to avoid compact groups or cluster environments. This gives us 146 isolated M\,83 analogs.
We then mock-observe these analogs by putting them at a distance of 4.9\,Mpc with a random orientation. We exclude all subhalos falling within 0.2 degree -- the optical diameter of M\,83, obfuscating any potential dwarf galaxy lying in this region. We constructed  satellite catalogs where we select all subhalos within 330\,kpc, which corresponds to farthest projected M\,83 satellite distance, and within a depth of 350\,kpc along the line-of-sight. This latter value was chosen to correspond to the observed line-of-sight depth in our sample, corrected for the typical distance uncertainties, and ensures that no overlap with the satellite system of another nearby host is to be expected given our 700 kpc isolation criterion. 

  The results of our comparison of the luminosity function to Illustris-TNG50 are shown in Fig.\,\ref{fig:illustris}. For three of the analogs, we did not observe any satellite brighter than $M_V=-10$ whatsoever. Another two have only one such satellite. Most of the simulated analogs have several satellite galaxies within the detection volume, but overall their number is smaller than that of the confirmed observed satellite galaxies. When compared to the observations, this means that the number of predicted subhalos in the simulation is typically lower than the observed luminosity function. 
However, there are still a few M\,83 analogs at higher masses having similar luminosity functions at the bright end ($M_V<-15$).  In Fig.\,\ref{fig:illustris} we provide the percentage of analogs having a higher abundance of satellites in the simulation than observed, given at the luminosity steps of every observed satellite. {At the bright end, the observed and expected luminosity functions agree well within 1$\sigma$. However, this quickly changes at the faint end.} The  discrepancy is largest at M$_V=-14$\,mag, which is in tension with the simulated analogs by over 3$\sigma$. 
 This is intriguing, because the observed luminosity function is based on a shallow dwarf galaxy survey \citep{2015A&A...583A..79M}, meaning that the abundance of dwarf galaxies {in this part of the luminosity function} is likely underestimated --  
 some dwarf galaxies of M\,83 may be hidden. \citet{2015A&A...583A..79M} injected artificial galaxies  in the survey data and found that even at the bright end of the luminosity function, only 80 percent of the artificial galaxies were rediscovered. This is due to the high density of stars and contamination of Galactic cirrus patches, making the detection of potential dwarf galaxies a hard task. This means that in our work here we even may underestimate the real discrepancy between the number of observed satellites compared to the predicted abundance. Including this incompleteness in the comparison will move the predicted mean luminosity function and its bounds down by 20\%, which increases the tension from 3 to 4$\sigma$.

\subsection{Sub-halo mass function}

\begin{figure}
\centering
\includegraphics[width=\linewidth]{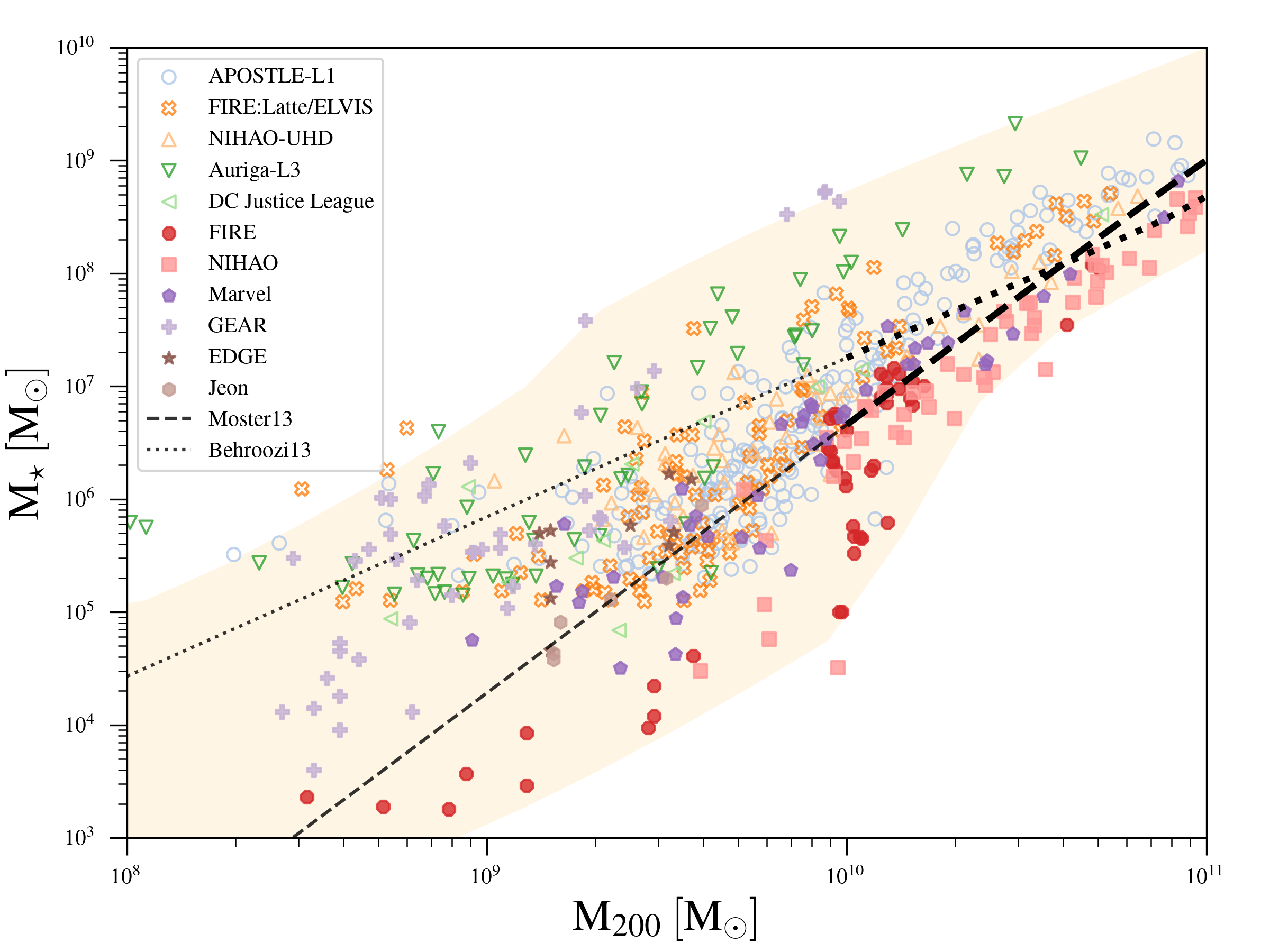}
 \caption{\label{fig:subhalofunctio} The stellar to halo mass function adapted from Fig.\,2 of \citet{2022NatAs...6..897S}. The orange area indicates the area used in the sampling of the luminosity function.
 Open symbols correspond to simulations of dwarf galaxies, 
 around a MW-mass host halo from:
 APOSTLE-l1 \citep{sawala2016,fattahi2016},
 FIRE:Latte/ELVIS \citep{wetzel2016,garrison-kimmel2019},
 NIHAO-UHD \citep{buck2019},
 Auriga-L3 \citep{grand2017},
 DC Justice League \citep{munshi2021}.
 Filled symbols correspond to zoom-in dwarf galaxies from:
 FIRE \citep{wheeler2015,fitts2017,wheeler2019},
 NIHAO \citep{wang2015},
 Marvel \citep{munshi2021},
 GEAR \citep{2018A&A...616A..96R,sanati2023},
 EDGE \citep{rey2019,rey2020},
 Jeon \citep{jeon2017}.
 The dashed and dotted lines show abundance matching relations from \citep{behroozi2013} and \citep{moster2013} respectively. Note that those predictions are extrapolated below about $10^{10}\rm{M_{200}}$. 
 }
\end{figure}

\begin{figure}
  \begin{center}
  \includegraphics[width=\linewidth]{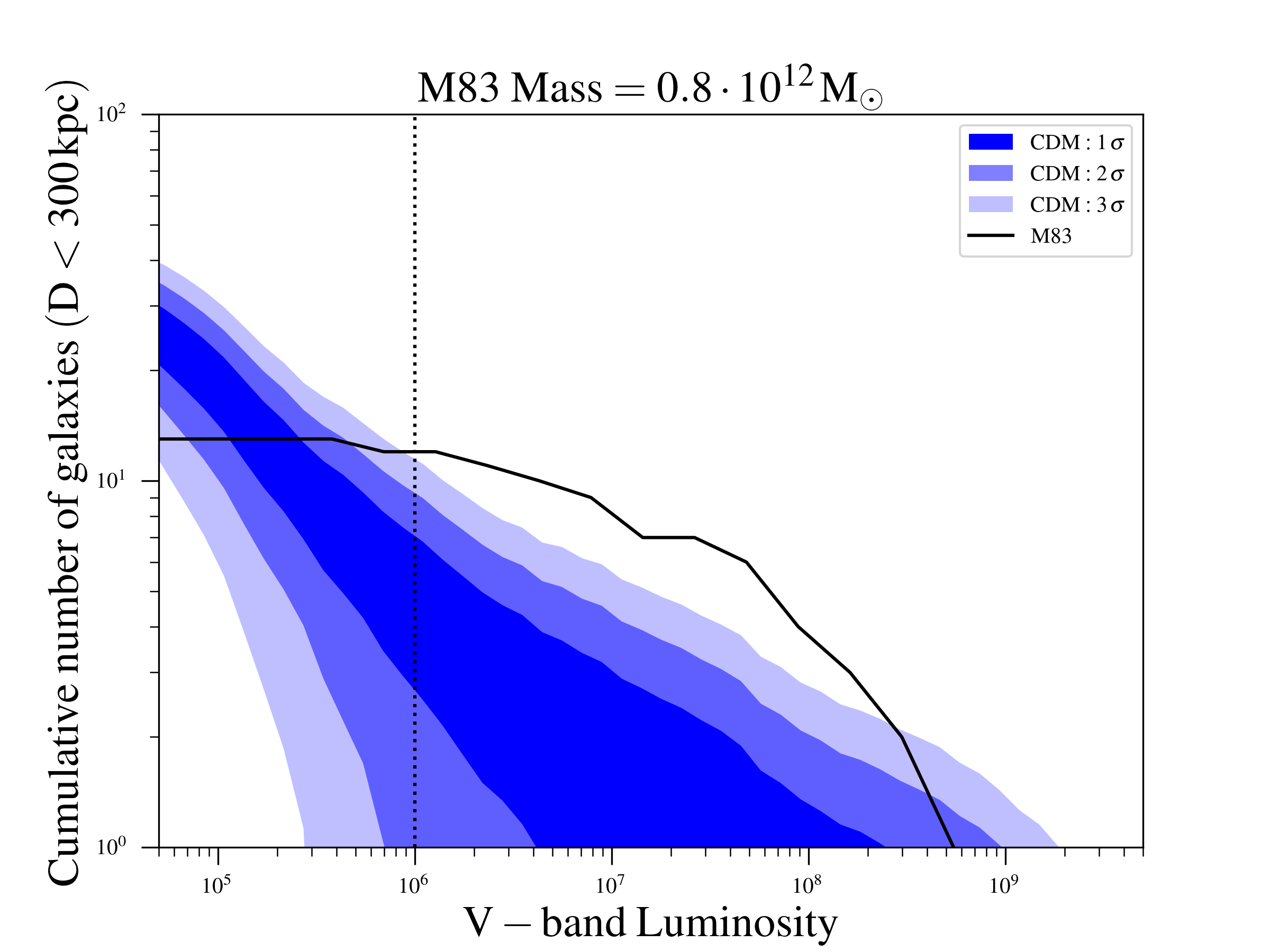}
  \caption{Predicted cumulative number of satellites brighter than
  a given luminosity, within 300 kpc around M83, assuming a mass of $0.8\times10^{12}\rm{M_{\odot}}$. The 1, 2, and 3$\sigma$ confidence intervals are indicated with the blue areas. The black line corresponds to the luminosity function of M\,83. The vertical dashed line represents the survey limit of \citet{2015A&A...583A..79M}.  The M\,83 luminosity function deviates by more than 3$\sigma$ from the sub-halo mass function.}
  \label{LCum}
  \end{center}
\end{figure}

Is this finding dependent on the cosmological simulation we use for our comparison? To check the robustness of our result, we now follow a different approach. Instead of counting the number of simulated dwarf galaxies we use the sub-halo mass function (i.e. the mass distribution of dark halos potentially hosting satellites around a host galaxy), providing the number of dark halos per mass unit for a given mass of the host galaxy  \citep{2015MNRAS.451.3117S}.
To estimate the satellite population around host galaxies we attributed a stellar mass to each sub-halo obtained (i.e. painting the dark matter sub-halos with dwarf galaxies). Contrary to Milky Way-like galaxies, dwarf galaxies
do not follow a well defined stellar mass - halo mass relation ($M_{\star} - M_{\rm sh}$). On the contrary, as seen on Fig.~2 of \citet{2022NatAs...6..897S},
a large scatter exists for a given halo mass, showing a variation of up to three dex in luminosity for a given halo mass. This scatter reflects different build-up histories of dwarf galaxies, which is further increased based on different models of galaxy formation yielding different results. 
To convert halo mass to stellar mass, we defined  based on Fig.~2  of \citet{2022NatAs...6..897S} a region in the $M_{\star} - M_{\rm sh}$ plot that encompasses all galaxies, either including only galaxies formed in the field  or satellite galaxies (i.e. combining their Fig.~2a and ~2b). {In Fig.\,\ref{fig:subhalofunctio} we combine their  Fig.~2a and ~2b.} We choose the area such that it contains the largest number of luminous subhalos (i.e. potentially over-predicting the number of dwarf galaxies).
For each sub-halo we randomly determine a luminosity
from a uniform distribution spanning the minimal and maximal luminosity in this area for a given halo mass (corresponding to the shaded area in Fig.\,\ref{fig:subhalofunctio}). {As an example: a subhalo with a halo mass of $10^{9}$\,M$_\odot$ can be asigned a stellar mass between $\sim$$10^3$\,M$_\odot$ and $\sim$$5\times10^6$\,M$_\odot$, which corresponds to the lower and upper value y-values of the area at the x-value of the halo mass in Fig.\,\ref{fig:subhalofunctio}.}
We derive the luminosity function of satellites 
as the cumulative number of satellites observed around a given galaxy.
The scatter is estimated by 1, 2 and 3 standard deviations around the mean, obtained by averaging 300 realisations. For the halo mass of M\,83 we take $0.8\times10^{12}$ solar masses \citep{2007AJ....133..504K}. {These bounds are shown in Fig.\,\ref{LCum}, together with the observed luminosity function of M\,83.}

{At the bright end (i.e. $>$$5\times10^8$\,M$_\odot$), the observed luminosity function of M\,83 is consistent with the predicted range of luminosity functions at a 2$\sigma$ level. However, this quickly starts to change at lower luminosities. At its maximum discrepency -- this is between 10$^7$ to 10$^8$\,M$_\odot$ --} the observed luminosity function of M\,83
 is more than 5.5$\sigma$ away from the predictions given by this process of assigning dwarf galaxies to dark matter sub-halos. This discrepancy is consistent with our previous finding in Illustris-TNG50 and puts the M\,83 group at odds with $\Lambda$CDM. 
 Increasing the halo mass to $1.0\times10^{12}$ (i.e. the Milky Way halo mass estimate) will still result in a 3$\sigma$ discrepancy and does not alleviate the problem. Furthermore, this comparison again does not take into account that the observed luminosity function may be underestimated due to incomplete observations by ~20\%. Including such a correction would only increase the tension.

\section{Summary and conclusion}
To study the luminosity function of the M\,83 group, we constructed a catalog of dwarf galaxies and employed distance measurements to confirm or reject eight dwarf galaxy candidate members from the literature. We could reject six of them, while two are too faint to get a good distance estimate. To be conservative, we assume them to be background objects. This leaves us with 13 confirmed dwarf galaxies of the M\,83 group.
We have compared this dwarf galaxy sample to cosmological simulations, namely the Illustris-TNG50 project, as well as theoretical predictions from the sub-halo mass function. By applying observational constraints to the simulated M\,83 analogs in Illustris-TNG50, we derived the expected luminosity functions. {At the bright end, the luminosity functions agree with the observed one. However, at the faint end, this is,}  in the luminosity range between 10$^6$ to 10$^8$ solar masses (i.e. -10 to -14\,mag in the $V$ band) all simulated M\,83 analogs contained fewer satellites, and with 13 observed satellites within 330\,kpc the expectation was overshot by 3$\sigma$. The same is true when we used the sub-halo mass function to predict the abundance of dwarf galaxies, significantly underestimating the luminosity function at the 5.5$\sigma$ level for all galaxy formation models.  
Observational incompleteness may further increase this tension by $\sim$1$\sigma$, because even at the bright end, dwarf galaxies may not have been observationally detected due to the large density of foreground stars and contamination with galactic cirrus. The flattening of the observed luminosity function at 10$^6$ solar masses indicates further incompleteness, as we expect it would grow with a power law.

Modern cosmological simulations have been reported to more closely match the luminosity function of the Milky Way, apparently resolving the classical missing-satellite problem. Yet, our results demonstrate that in the M\,83 system, the number of satellites is underestimated at more than a 3$\sigma$ level by such models, indicating that there is rather a too-many satellite problem for the  M\,83 group. This implies that $\Lambda$CDM models do not seem to correctly predict the abundances (and physical properties) of dwarf galaxies for the full range of observed host galaxies in the nearby universe. {In an extension of the presented work here, Kanehisa et al. (submitted) draw a similar conclusion at a higher significance level based on data from the MATLAS survey.}
 These results should serve as a cautionary tale, highlighting the need to remain vigilant and to continue testing and improving the models even when they can explain the observations in our own Local Group.

\begin{acknowledgements}
{We thank the referee for the constructive report, which helped to clarify and improve the manuscript.}
O.~M. is grateful to the Swiss National Science Foundation for financial support under the grant numbers P400P2\_191123 and PZ00P2\_202104. M.~S.~P.~acknowledges funding of a Leibniz-Junior Research Group (project number J94/2020). The authors thank Scott Carlsten for the help provided on the measurement of the surface brightness fluctuation. The authors further thank Azadeh Fattahi for providing us with the data points used in Fig.~\ref{fig:subhalofunctio}.
\end{acknowledgements}

\bibliographystyle{aa}
\bibliography{aanda}

\begin{thebibliography}{109}
\expandafter\ifx\csname natexlab\endcsname\relax\def\natexlab#1{#1}\fi

\bibitem[{{Abazajian} {et~al.}(2003){Abazajian}, {Adelman-McCarthy},
  {Ag{\"u}eros}, {Allam}, {Anderson}, {Annis}, {Bahcall}, {Baldry}, {Bastian},
  {Berlind}, {Bernardi}, {Blanton}, {Blythe}, {Bochanski}, {Boroski},
  {Brewington}, {Briggs}, {Brinkmann}, {Brunner}, {Budav{\'a}ri}, {Carey},
  {Carr}, {Castander}, {Chiu}, {Collinge}, {Connolly}, {Covey}, {Csabai},
  {Dalcanton}, {Dodelson}, {Doi}, {Dong}, {Eisenstein}, {Evans}, {Fan},
  {Feldman}, {Finkbeiner}, {Friedman}, {Frieman}, {Fukugita}, {Gal},
  {Gillespie}, {Glazebrook}, {Gonzalez}, {Gray}, {Grebel}, {Grodnicki}, {Gunn},
  {Gurbani}, {Hall}, {Hao}, {Harbeck}, {Harris}, {Harris}, {Harvanek},
  {Hawley}, {Heckman}, {Helmboldt}, {Hendry}, {Hennessy}, {Hindsley}, {Hogg},
  {Holmgren}, {Holtzman}, {Homer}, {Hui}, {Ichikawa}, {Ichikawa}, {Inkmann},
  {Ivezi{\'c}}, {Jester}, {Johnston}, {Jordan}, {Jordan}, {Jorgensen},
  {Juri{\'c}}, {Kauffmann}, {Kent}, {Kleinman}, {Knapp}, {Kniazev}, {Kron},
  {Krzesi{\'n}ski}, {Kunszt}, {Kuropatkin}, {Lamb}, {Lampeitl}, {Laubscher},
  {Lee}, {Leger}, {Li}, {Lidz}, {Lin}, {Loh}, {Long}, {Loveday}, {Lupton},
  {Malik}, {Margon}, {McGehee}, {McKay}, {Meiksin}, {Miknaitis}, {Moorthy},
  {Munn}, {Murphy}, {Nakajima}, {Narayanan}, {Nash}, {Neilsen}, {Newberg},
  {Newman}, {Nichol}, {Nicinski}, {Nieto-Santisteban}, {Nitta}, {Odenkirchen},
  {Okamura}, {Ostriker}, {Owen}, {Padmanabhan}, {Peoples}, {Pier}, {Pindor},
  {Pope}, {Quinn}, {Rafikov}, {Raymond}, {Richards}, {Richmond}, {Rix},
  {Rockosi}, {Schaye}, {Schlegel}, {Schneider}, {Schroeder}, {Scranton},
  {Sekiguchi}, {Seljak}, {Sergey}, {Sesar}, {Sheldon}, {Shimasaku}, {Siegmund},
  {Silvestri}, {Sinisgalli}, {Sirko}, {Smith}, {Smol{\v{c}}i{\'c}}, {Snedden},
  {Stebbins}, {Steinhardt}, {Stinson}, {Stoughton}, {Strateva}, {Strauss},
  {SubbaRao}, {Szalay}, {Szapudi}, {Szkody}, {Tasca}, {Tegmark}, {Thakar},
  {Tremonti}, {Tucker}, {Uomoto}, {Vanden Berk}, {Vandenberg}, {Vogeley},
  {Voges}, {Vogt}, {Walkowicz}, {Weinberg}, {West}, {White}, {Wilhite},
  {Willman}, {Xu}, {Yanny}, {Yarger}, {Yasuda}, {Yip}, {Yocum}, {York},
  {Zakamska}, {Zehavi}, {Zheng}, {Zibetti}, \& {Zucker}}]{2003AJ....126.2081A}
{Abazajian}, K., {Adelman-McCarthy}, J.~K., {Ag{\"u}eros}, M.~A., {et~al.}
  2003, \aj, 126, 2081

\bibitem[{{Abraham} \& {van Dokkum}(2014)}]{2014PASP..126...55A}
{Abraham}, R.~G. \& {van Dokkum}, P.~G. 2014, \pasp, 126, 55

\bibitem[{{Bacon} {et~al.}(2010){Bacon}, {Accardo}, {Adjali}, {Anwand},
  {Bauer}, {Biswas}, {Blaizot}, {Boudon}, {Brau-Nogue}, {Brinchmann},
  {Caillier}, {Capoani}, {Carollo}, {Contini}, {Couderc}, {Daguis{\'e}},
  {Deiries}, {Delabre}, {Dreizler}, {Dubois}, {Dupieux}, {Dupuy}, {Emsellem},
  {Fechner}, {Fleischmann}, {Fran{\c{c}}ois}, {Gallou}, {Gharsa}, {Glindemann},
  {Gojak}, {Guiderdoni}, {Hansali}, {Hahn}, {Jarno}, {Kelz}, {Koehler},
  {Kosmalski}, {Laurent}, {Le Floch}, {Lilly}, {Lizon}, {Loupias}, {Manescau},
  {Monstein}, {Nicklas}, {Olaya}, {Pares}, {Pasquini}, {P{\'e}contal-Rousset},
  {Pell{\'o}}, {Petit}, {Popow}, {Reiss}, {Remillieux}, {Renault}, {Roth},
  {Rupprecht}, {Serre}, {Schaye}, {Soucail}, {Steinmetz}, {Streicher}, {Stuik},
  {Valentin}, {Vernet}, {Weilbacher}, {Wisotzki}, \&
  {Yerle}}]{2010SPIE.7735E..08B}
{Bacon}, R., {Accardo}, M., {Adjali}, L., {et~al.} 2010, in Society of
  Photo-Optical Instrumentation Engineers (SPIE) Conference Series, Vol. 7735,
  Ground-based and Airborne Instrumentation for Astronomy III, ed. I.~S.
  {McLean}, S.~K. {Ramsay}, \& H.~{Takami}, 773508

\bibitem[{{Banks} {et~al.}(1999){Banks}, {Disney}, {Knezek}, {Jerjen},
  {Barnes}, {Bhatal}, {de Blok}, {Boyce}, {Ekers}, {Freeman}, {Gibson},
  {Henning}, {Kilborn}, {Koribalski}, {Kraan-Korteweg}, {Malin}, {Minchin},
  {Mould}, {Oosterloo}, {Price}, {Putman}, {Ryder}, {Sadler}, {Staveley-Smith},
  {Stewart}, {Stootman}, {Vaile}, {Webster}, \& {Wright}}]{1999ApJ...524..612B}
{Banks}, G.~D., {Disney}, M.~J., {Knezek}, P.~M., {et~al.} 1999, \apj, 524, 612

\bibitem[{{Barnes} {et~al.}(2014){Barnes}, {van Zee}, {Dale}, {Staudaher},
  {Bullock}, {Calzetti}, {Chandar}, \& {Dalcanton}}]{2014ApJ...789..126B}
{Barnes}, K.~L., {van Zee}, L., {Dale}, D.~A., {et~al.} 2014, \apj, 789, 126

\bibitem[{{Beasley} {et~al.}(2023){Beasley}, {Fahrion}, \&
  {Gvozdenko}}]{2023arXiv231201420B}
{Beasley}, M.~A., {Fahrion}, K., \& {Gvozdenko}, A. 2023, arXiv e-prints,
  arXiv:2312.01420

\bibitem[{{Begum} {et~al.}(2008){Begum}, {Chengalur}, {Karachentsev},
  {Sharina}, \& {Kaisin}}]{2008MNRAS.386.1667B}
{Begum}, A., {Chengalur}, J.~N., {Karachentsev}, I.~D., {Sharina}, M.~E., \&
  {Kaisin}, S.~S. 2008, \mnras, 386, 1667

\bibitem[{{Behroozi} {et~al.}(2013){Behroozi}, {Wechsler}, \&
  {Conroy}}]{behroozi2013}
{Behroozi}, P.~S., {Wechsler}, R.~H., \& {Conroy}, C. 2013, \apj, 770, 57

\bibitem[{{Bennet} {et~al.}(2017){Bennet}, {Sand}, {Crnojevi{\'c}}, {Spekkens},
  {Zaritsky}, \& {Karunakaran}}]{2017ApJ...850..109B}
{Bennet}, P., {Sand}, D.~J., {Crnojevi{\'c}}, D., {et~al.} 2017, \apj, 850, 109

\bibitem[{{Bhardwaj}(2020)}]{2020JApA...41...23B}
{Bhardwaj}, A. 2020, Journal of Astrophysics and Astronomy, 41, 23

\bibitem[{{Bird} {et~al.}(2015){Bird}, {Flynn}, {Harris}, \&
  {Valtonen}}]{2015A&A...575A..72B}
{Bird}, S.~A., {Flynn}, C., {Harris}, W.~E., \& {Valtonen}, M. 2015, \aap, 575,
  A72

\bibitem[{{Boylan-Kolchin} {et~al.}(2011){Boylan-Kolchin}, {Bullock}, \&
  {Kaplinghat}}]{2011MNRAS.415L..40B}
{Boylan-Kolchin}, M., {Bullock}, J.~S., \& {Kaplinghat}, M. 2011, \mnras, 415,
  L40

\bibitem[{{Buck} {et~al.}(2019){Buck}, {Macci{\`o}}, {Dutton}, {Obreja}, \&
  {Frings}}]{buck2019}
{Buck}, T., {Macci{\`o}}, A.~V., {Dutton}, A.~A., {Obreja}, A., \& {Frings}, J.
  2019, \mnras, 483, 1314

\bibitem[{{Bull} {et~al.}(2016){Bull}, {Akrami}, {Adamek}, {Baker}, {Bellini},
  {Beltr{\'a}n Jim{\'e}nez}, {Bentivegna}, {Camera}, {Clesse}, {Davis}, {Di
  Dio}, {Enander}, {Heavens}, {Heisenberg}, {Hu}, {Llinares}, {Maartens},
  {M{\"o}rtsell}, {Nadathur}, {Noller}, {Pasechnik}, {Pawlowski}, {Pereira},
  {Quartin}, {Ricciardone}, {Riemer-S{\o}rensen}, {Rinaldi}, {Sakstein},
  {Saltas}, {Salzano}, {Sawicki}, {Solomon}, {Spolyar}, {Starkman}, {Steer},
  {Tereno}, {Verde}, {Villaescusa-Navarro}, {von Strauss}, \&
  {Winther}}]{2016PDU....12...56B}
{Bull}, P., {Akrami}, Y., {Adamek}, J., {et~al.} 2016, Physics of the Dark
  Universe, 12, 56

\bibitem[{{Bullock} \& {Boylan-Kolchin}(2017)}]{Bullock2017}
{Bullock}, J.~S. \& {Boylan-Kolchin}, M. 2017, \araa, 55, 343

\bibitem[{{Carlsten} {et~al.}(2019{\natexlab{a}}){Carlsten}, {Beaton}, {Greco},
  \& {Greene}}]{2019ApJ...879...13C}
{Carlsten}, S.~G., {Beaton}, R.~L., {Greco}, J.~P., \& {Greene}, J.~E.
  2019{\natexlab{a}}, \apj, 879, 13

\bibitem[{{Carlsten} {et~al.}(2019{\natexlab{b}}){Carlsten}, {Beaton}, {Greco},
  \& {Greene}}]{2019ApJ...878L..16C}
{Carlsten}, S.~G., {Beaton}, R.~L., {Greco}, J.~P., \& {Greene}, J.~E.
  2019{\natexlab{b}}, \apjl, 878, L16

\bibitem[{{Carlsten} {et~al.}(2022){Carlsten}, {Greene}, {Beaton}, {Danieli},
  \& {Greco}}]{2022ApJ...933...47C}
{Carlsten}, S.~G., {Greene}, J.~E., {Beaton}, R.~L., {Danieli}, S., \& {Greco},
  J.~P. 2022, \apj, 933, 47

\bibitem[{{Carlsten} {et~al.}(2021){Carlsten}, {Greene}, {Peter}, {Beaton}, \&
  {Greco}}]{2021ApJ...908..109C}
{Carlsten}, S.~G., {Greene}, J.~E., {Peter}, A. H.~G., {Beaton}, R.~L., \&
  {Greco}, J.~P. 2021, \apj, 908, 109

\bibitem[{{Carrillo} {et~al.}(2017){Carrillo}, {Bell}, {Bailin}, {Monachesi},
  {de Jong}, {Harmsen}, \& {Slater}}]{2017MNRAS.465.5026C}
{Carrillo}, A., {Bell}, E.~F., {Bailin}, J., {et~al.} 2017, \mnras, 465, 5026

\bibitem[{{Cautun} {et~al.}(2020){Cautun}, {Ben{\'\i}tez-Llambay}, {Deason},
  {Frenk}, {Fattahi}, {G{\'o}mez}, {Grand}, {Oman}, {Navarro}, \&
  {Simpson}}]{2020MNRAS.494.4291C}
{Cautun}, M., {Ben{\'\i}tez-Llambay}, A., {Deason}, A.~J., {et~al.} 2020,
  \mnras, 494, 4291

\bibitem[{{Chiboucas} {et~al.}(2013){Chiboucas}, {Jacobs}, {Tully}, \&
  {Karachentsev}}]{2013AJ....146..126C}
{Chiboucas}, K., {Jacobs}, B.~A., {Tully}, R.~B., \& {Karachentsev}, I.~D.
  2013, \aj, 146, 126

\bibitem[{{Chiboucas} {et~al.}(2009){Chiboucas}, {Karachentsev}, \&
  {Tully}}]{2009AJ....137.3009C}
{Chiboucas}, K., {Karachentsev}, I.~D., \& {Tully}, R.~B. 2009, \aj, 137, 3009

\bibitem[{{Crnojevi{\'c}} {et~al.}(2012){Crnojevi{\'c}}, {Grebel}, \&
  {Cole}}]{2012A&A...541A.131C}
{Crnojevi{\'c}}, D., {Grebel}, E.~K., \& {Cole}, A.~A. 2012, \aap, 541, A131

\bibitem[{{Crnojevi{\'c}} {et~al.}(2019){Crnojevi{\'c}}, {Sand}, {Bennet},
  {Pasetto}, {Spekkens}, {Caldwell}, {Guhathakurta}, {McLeod}, {Seth}, {Simon},
  {Strader}, \& {Toloba}}]{2019ApJ...872...80C}
{Crnojevi{\'c}}, D., {Sand}, D.~J., {Bennet}, P., {et~al.} 2019, \apj, 872, 80

\bibitem[{{Crnojevi{\'c}} {et~al.}(2014){Crnojevi{\'c}}, {Sand}, {Caldwell},
  {Guhathakurta}, {McLeod}, {Seth}, {Simon}, {Strader}, \&
  {Toloba}}]{2014ApJ...795L..35C}
{Crnojevi{\'c}}, D., {Sand}, D.~J., {Caldwell}, N., {et~al.} 2014, \apjl, 795,
  L35

\bibitem[{{Crnojevi{\'c}} {et~al.}(2016){Crnojevi{\'c}}, {Sand}, {Spekkens},
  {Caldwell}, {Guhathakurta}, {McLeod}, {Seth}, {Simon}, {Strader}, \&
  {Toloba}}]{2016ApJ...823...19C}
{Crnojevi{\'c}}, D., {Sand}, D.~J., {Spekkens}, K., {et~al.} 2016, \apj, 823,
  19

\bibitem[{{Crosby} {et~al.}(2023{\natexlab{a}}){Crosby}, {Jerjen},
  {M{\"u}ller}, {Pawlowski}, {Mateo}, \& {Dirnberger}}]{2023MNRAS.521.4009C}
{Crosby}, E., {Jerjen}, H., {M{\"u}ller}, O., {et~al.} 2023{\natexlab{a}},
  \mnras, 521, 4009

\bibitem[{{Crosby} {et~al.}(2023{\natexlab{b}}){Crosby}, {Jerjen},
  {M{\"u}ller}, {Pawlowski}, {Mateo}, \& {Lelli}}]{2023MNRAS.tmp.3597C}
{Crosby}, E., {Jerjen}, H., {M{\"u}ller}, O., {et~al.} 2023{\natexlab{b}},
  \mnras [\eprint[arXiv]{2312.03486}]

\bibitem[{{Da Costa} \& {Armandroff}(1990)}]{daCostaArmandroff1990}
{Da Costa}, G.~S. \& {Armandroff}, T.~E. 1990, \aj, 100, 162

\bibitem[{{Da Costa} {et~al.}(2010){Da Costa}, {Rejkuba}, {Jerjen}, \&
  {Grebel}}]{2010ApJ...708L.121D}
{Da Costa}, G.~S., {Rejkuba}, M., {Jerjen}, H., \& {Grebel}, E.~K. 2010, \apj,
  708, L121

\bibitem[{{Dey} {et~al.}(2019){Dey}, {Schlegel}, {Lang}, {Blum}, {Burleigh},
  {Fan}, {Findlay}, {Finkbeiner}, {Herrera}, {Juneau}, {Landriau}, {Levi},
  {McGreer}, {Meisner}, {Myers}, {Moustakas}, {Nugent}, {Patej}, {Schlafly},
  {Walker}, {Valdes}, {Weaver}, {Y{\`e}che}, {Zou}, {Zhou}, {Abareshi},
  {Abbott}, {Abolfathi}, {Aguilera}, {Alam}, {Allen}, {Alvarez}, {Annis},
  {Ansarinejad}, {Aubert}, {Beechert}, {Bell}, {BenZvi}, {Beutler}, {Bielby},
  {Bolton}, {Brice{\~n}o}, {Buckley-Geer}, {Butler}, {Calamida}, {Carlberg},
  {Carter}, {Casas}, {Castander}, {Choi}, {Comparat}, {Cukanovaite}, {Delubac},
  {DeVries}, {Dey}, {Dhungana}, {Dickinson}, {Ding}, {Donaldson}, {Duan},
  {Duckworth}, {Eftekharzadeh}, {Eisenstein}, {Etourneau}, {Fagrelius},
  {Farihi}, {Fitzpatrick}, {Font-Ribera}, {Fulmer}, {G{\"a}nsicke},
  {Gaztanaga}, {George}, {Gerdes}, {Gontcho}, {Gorgoni}, {Green}, {Guy},
  {Harmer}, {Hernandez}, {Honscheid}, {Huang}, {James}, {Jannuzi}, {Jiang},
  {Joyce}, {Karcher}, {Karkar}, {Kehoe}, {Kneib}, {Kueter-Young}, {Lan},
  {Lauer}, {Le Guillou}, {Le Van Suu}, {Lee}, {Lesser}, {Perreault Levasseur},
  {Li}, {Mann}, {Marshall}, {Mart{\'\i}nez-V{\'a}zquez}, {Martini}, {du Mas des
  Bourboux}, {McManus}, {Meier}, {M{\'e}nard}, {Metcalfe},
  {Mu{\~n}oz-Guti{\'e}rrez}, {Najita}, {Napier}, {Narayan}, {Newman}, {Nie},
  {Nord}, {Norman}, {Olsen}, {Paat}, {Palanque-Delabrouille}, {Peng},
  {Poppett}, {Poremba}, {Prakash}, {Rabinowitz}, {Raichoor}, {Rezaie},
  {Robertson}, {Roe}, {Ross}, {Ross}, {Rudnick}, {Safonova}, {Saha},
  {S{\'a}nchez}, {Savary}, {Schweiker}, {Scott}, {Seo}, {Shan}, {Silva},
  {Slepian}, {Soto}, {Sprayberry}, {Staten}, {Stillman}, {Stupak}, {Summers},
  {Sien Tie}, {Tirado}, {Vargas-Maga{\~n}a}, {Vivas}, {Wechsler}, {Williams},
  {Yang}, {Yang}, {Yapici}, {Zaritsky}, {Zenteno}, {Zhang}, {Zhang}, {Zhou}, \&
  {Zhou}}]{2019AJ....157..168D}
{Dey}, A., {Schlegel}, D.~J., {Lang}, D., {et~al.} 2019, \aj, 157, 168

\bibitem[{{Duc} {et~al.}(2015){Duc}, {Cuillandre}, {Karabal}, {Cappellari},
  {Alatalo}, {Blitz}, {Bournaud}, {Bureau}, {Crocker}, {Davies}, {Davis}, {de
  Zeeuw}, {Emsellem}, {Khochfar}, {Krajnovi{\'c}}, {Kuntschner}, {McDermid},
  {Michel-Dansac}, {Morganti}, {Naab}, {Oosterloo}, {Paudel}, {Sarzi}, {Scott},
  {Serra}, {Weijmans}, \& {Young}}]{2015MNRAS.446..120D}
{Duc}, P.-A., {Cuillandre}, J.-C., {Karabal}, E., {et~al.} 2015, \mnras, 446,
  120

\bibitem[{{Dykes} {et~al.}(2021){Dykes}, {Gheller}, {Koribalski}, {Dolag}, \&
  {Krokos}}]{2021A&C....3400448D}
{Dykes}, T., {Gheller}, C., {Koribalski}, B.~S., {Dolag}, K., \& {Krokos}, M.
  2021, Astronomy and Computing, 34, 100448

\bibitem[{{Fattahi} {et~al.}(2016){Fattahi}, {Navarro}, {Sawala}, {Frenk},
  {Oman}, {Crain}, {Furlong}, {Schaller}, {Schaye}, {Theuns}, \&
  {Jenkins}}]{fattahi2016}
{Fattahi}, A., {Navarro}, J.~F., {Sawala}, T., {et~al.} 2016, \mnras, 457, 844

\bibitem[{{Fitts} {et~al.}(2017){Fitts}, {Boylan-Kolchin}, {Elbert}, {Bullock},
  {Hopkins}, {O{\~n}orbe}, {Wetzel}, {Wheeler}, {Faucher-Gigu{\`e}re},
  {Kere{\v{s}}}, {Skillman}, \& {Weisz}}]{fitts2017}
{Fitts}, A., {Boylan-Kolchin}, M., {Elbert}, O.~D., {et~al.} 2017, \mnras, 471,
  3547

\bibitem[{{For} {et~al.}(2019){For}, {Staveley-Smith}, {Westmeier}, {Whiting},
  {Oh}, {Koribalski}, {Wang}, {Wong}, {Bekiaris}, {Cortese}, {Elagali},
  {Kleiner}, {Lee-Waddell}, {Madrid}, {Popping}, {Rhee}, {Reynolds}, {Collier},
  {Phillips}, {Voronkov}, {M{\"u}ller}, \& {Jerjen}}]{2019MNRAS.489.5723F}
{For}, B.~Q., {Staveley-Smith}, L., {Westmeier}, T., {et~al.} 2019, \mnras,
  489, 5723

\bibitem[{{Garrison-Kimmel} {et~al.}(2014){Garrison-Kimmel}, {Boylan-Kolchin},
  {Bullock}, \& {Kirby}}]{2014MNRAS.444..222G}
{Garrison-Kimmel}, S., {Boylan-Kolchin}, M., {Bullock}, J.~S., \& {Kirby},
  E.~N. 2014, \mnras, 444, 222

\bibitem[{{Garrison-Kimmel} {et~al.}(2019){Garrison-Kimmel}, {Hopkins},
  {Wetzel}, {Bullock}, {Boylan-Kolchin}, {Kere{\v{s}}}, {Faucher-Gigu{\`e}re},
  {El-Badry}, {Lamberts}, {Quataert}, \& {Sanderson}}]{garrison-kimmel2019}
{Garrison-Kimmel}, S., {Hopkins}, P.~F., {Wetzel}, A., {et~al.} 2019, \mnras,
  487, 1380

\bibitem[{{Grand} {et~al.}(2017){Grand}, {G{\'o}mez}, {Marinacci}, {Pakmor},
  {Springel}, {Campbell}, {Frenk}, {Jenkins}, \& {White}}]{grand2017}
{Grand}, R. J.~J., {G{\'o}mez}, F.~A., {Marinacci}, F., {et~al.} 2017, \mnras,
  467, 179

\bibitem[{{Grossi} {et~al.}(2007){Grossi}, {Disney}, {Pritzl}, {Knezek},
  {Gallagher}, {Minchin}, \& {Freeman}}]{2007MNRAS.374..107G}
{Grossi}, M., {Disney}, M.~J., {Pritzl}, B.~J., {et~al.} 2007, \mnras, 374, 107

\bibitem[{{Gwyn}(2008)}]{2008PASP..120..212G}
{Gwyn}, S. D.~J. 2008, \pasp, 120, 212

\bibitem[{{Habas} {et~al.}(2020){Habas}, {Marleau}, {Duc}, {Durrell}, {Paudel},
  {Poulain}, {S{\'a}nchez-Janssen}, {Sreejith}, {Ramasawmy}, {Stemock},
  {Leach}, {Cuillandre}, {Gwyn}, {Agnello}, {B{\'\i}lek}, {Fensch},
  {M{\"u}ller}, {Peng}, \& {van der Burg}}]{2020MNRAS.491.1901H}
{Habas}, R., {Marleau}, F.~R., {Duc}, P.-A., {et~al.} 2020, \mnras, 491, 1901

\bibitem[{{Helmi} {et~al.}(2018){Helmi}, {Babusiaux}, {Koppelman}, {Massari},
  {Veljanoski}, \& {Brown}}]{2018Natur.563...85H}
{Helmi}, A., {Babusiaux}, C., {Koppelman}, H.~H., {et~al.} 2018, \nat, 563, 85

\bibitem[{{Herrmann} {et~al.}(2008){Herrmann}, {Ciardullo}, {Feldmeier}, \&
  {Vinciguerra}}]{2008ApJ...683..630H}
{Herrmann}, K.~A., {Ciardullo}, R., {Feldmeier}, J.~J., \& {Vinciguerra}, M.
  2008, \apj, 683, 630

\bibitem[{{Huchtmeier} \& {Bohnenstengel}(1981)}]{1981A&A...100...72H}
{Huchtmeier}, W.~K. \& {Bohnenstengel}, H.~D. 1981, \aap, 100, 72

\bibitem[{{Ibata} {et~al.}(2013){Ibata}, {Lewis}, {Conn}, {Irwin},
  {McConnachie}, {Chapman}, {Collins}, {Fardal}, {Ferguson}, {Ibata}, {Mackey},
  {Martin}, {Navarro}, {Rich}, {Valls-Gabaud}, \&
  {Widrow}}]{2013Natur.493...62I}
{Ibata}, R.~A., {Lewis}, G.~F., {Conn}, A.~R., {et~al.} 2013, NAT, 493, 62

\bibitem[{{Javanmardi} {et~al.}(2016){Javanmardi}, {Martinez-Delgado},
  {Kroupa}, {Henkel}, {Crawford}, {Teuwen}, {Gabany}, {Hanson}, {Chonis}, \&
  {Neyer}}]{2016A&A...588A..89J}
{Javanmardi}, B., {Martinez-Delgado}, D., {Kroupa}, P., {et~al.} 2016, \aap,
  588, A89

\bibitem[{{Javanmardi} {et~al.}(2019){Javanmardi}, {Raouf}, {Khosroshahi},
  {Tavasoli}, {M{\"u}ller}, \& {Molaeinezhad}}]{2019ApJ...870...50J}
{Javanmardi}, B., {Raouf}, M., {Khosroshahi}, H.~G., {et~al.} 2019, \apj, 870,
  50

\bibitem[{{Jeon} {et~al.}(2017){Jeon}, {Besla}, \& {Bromm}}]{jeon2017}
{Jeon}, M., {Besla}, G., \& {Bromm}, V. 2017, \apj, 848, 85

\bibitem[{{Jerjen} \& {Rejkuba}(2001)}]{2001A&A...371..487J}
{Jerjen}, H. \& {Rejkuba}, M. 2001, \aap, 371, 487

\bibitem[{{Kamphuis} {et~al.}(2015){Kamphuis}, {J{\'o}zsa}, {Oh}, {Spekkens},
  {Urbancic}, {Serra}, {Koribalski}, \& {Dettmar}}]{2015MNRAS.452.3139K}
{Kamphuis}, P., {J{\'o}zsa}, G.~I.~G., {Oh}, S. .~H., {et~al.} 2015, \mnras,
  452, 3139

\bibitem[{{Kanehisa} {et~al.}(2023){Kanehisa}, {Pawlowski}, \&
  {M{\"u}ller}}]{2023MNRAS.524..952K}
{Kanehisa}, K.~J., {Pawlowski}, M.~S., \& {M{\"u}ller}, O. 2023, \mnras, 524,
  952

\bibitem[{{Karachentsev} {et~al.}(2013){Karachentsev}, {Makarov}, \&
  {Kaisina}}]{2013AJ....145..101K}
{Karachentsev}, I.~D., {Makarov}, D.~I., \& {Kaisina}, E.~I. 2013, \aj, 145,
  101

\bibitem[{{Karachentsev} {et~al.}(2008){Karachentsev}, {Makarov},
  {Karachentseva}, \& {Melnik}}]{2008AstL...34..832K}
{Karachentsev}, I.~D., {Makarov}, D.~I., {Karachentseva}, V.~E., \& {Melnik},
  O.~V. 2008, Astronomy Letters, 34, 832

\bibitem[{{Karachentsev} {et~al.}(2002){Karachentsev}, {Sharina}, {Dolphin},
  {Grebel}, {Geisler}, {Guhathakurta}, {Hodge}, {Karachentseva}, {Sarajedini},
  \& {Seitzer}}]{2002A&A...385...21K}
{Karachentsev}, I.~D., {Sharina}, M.~E., {Dolphin}, A.~E., {et~al.} 2002, \aap,
  385, 21

\bibitem[{{Karachentsev} {et~al.}(2007){Karachentsev}, {Tully}, {Dolphin},
  {Sharina}, {Makarova}, {Makarov}, {Sakai}, {Shaya}, {Kashibadze},
  {Karachentseva}, \& {Rizzi}}]{2007AJ....133..504K}
{Karachentsev}, I.~D., {Tully}, R.~B., {Dolphin}, A., {et~al.} 2007, \aj, 133,
  504

\bibitem[{{Koribalski} {et~al.}(2004){Koribalski}, {Staveley-Smith}, {Kilborn},
  {Ryder}, {Kraan-Korteweg}, {Ryan-Weber}, {Ekers}, {Jerjen}, {Henning},
  {Putman}, {Zwaan}, {de Blok}, {Calabretta}, {Disney}, {Minchin}, {Bhathal},
  {Boyce}, {Drinkwater}, {Freeman}, {Gibson}, {Green}, {Haynes}, {Juraszek},
  {Kesteven}, {Knezek}, {Mader}, {Marquarding}, {Meyer}, {Mould}, {Oosterloo},
  {O'Brien}, {Price}, {Sadler}, {Schr{\"o}der}, {Stewart}, {Stootman}, {Waugh},
  {Warren}, {Webster}, \& {Wright}}]{2004AJ....128...16K}
{Koribalski}, B.~S., {Staveley-Smith}, L., {Kilborn}, V.~A., {et~al.} 2004,
  \aj, 128, 16

\bibitem[{{Kroupa} {et~al.}(2010){Kroupa}, {Famaey}, {de Boer},
  {Dabringhausen}, {Pawlowski}, {Boily}, {Jerjen}, {Forbes}, {Hensler}, \&
  {Metz}}]{2010A&A...523A..32K}
{Kroupa}, P., {Famaey}, B., {de Boer}, K.~S., {et~al.} 2010, \aap, 523, A32

\bibitem[{{Leavitt} \& {Pickering}(1912)}]{1912HarCi.173....1L}
{Leavitt}, H.~S. \& {Pickering}, E.~C. 1912, Harvard College Observatory
  Circular, 173, 1

\bibitem[{{Lee} {et~al.}(1993){Lee}, {Freedman}, \& {Madore}}]{Lee1993}
{Lee}, M.~G., {Freedman}, W.~L., \& {Madore}, B.~F. 1993, \apj, 417, 553

\bibitem[{{Libeskind} {et~al.}(2015){Libeskind}, {Hoffman}, {Tully},
  {Courtois}, {Pomar{\`e}de}, {Gottl{\"o}ber}, \&
  {Steinmetz}}]{2015MNRAS.452.1052L}
{Libeskind}, N.~I., {Hoffman}, Y., {Tully}, R.~B., {et~al.} 2015, \mnras, 452,
  1052

\bibitem[{{McGaugh}(2005)}]{2005ApJ...632..859M}
{McGaugh}, S.~S. 2005, \apj, 632, 859

\bibitem[{{Meyer} {et~al.}(2004){Meyer}, {Zwaan}, {Webster}, {Staveley-Smith},
  {Ryan-Weber}, {Drinkwater}, {Barnes}, {Howlett}, {Kilborn}, {Stevens},
  {Waugh}, {Pierce}, {Bhathal}, {de Blok}, {Disney}, {Ekers}, {Freeman},
  {Garcia}, {Gibson}, {Harnett}, {Henning}, {Jerjen}, {Kesteven}, {Knezek},
  {Koribalski}, {Mader}, {Marquarding}, {Minchin}, {O'Brien}, {Oosterloo},
  {Price}, {Putman}, {Ryder}, {Sadler}, {Stewart}, {Stootman}, \&
  {Wright}}]{2004MNRAS.350.1195M}
{Meyer}, M.~J., {Zwaan}, M.~A., {Webster}, R.~L., {et~al.} 2004, \mnras, 350,
  1195

\bibitem[{{Miyazaki} {et~al.}(2018){Miyazaki}, {Komiyama}, {Kawanomoto}, {Doi},
  {Furusawa}, {Hamana}, {Hayashi}, {Ikeda}, {Kamata}, {Karoji}, {Koike},
  {Kurakami}, {Miyama}, {Morokuma}, {Nakata}, {Namikawa}, {Nakaya}, {Nariai},
  {Obuchi}, {Oishi}, {Okada}, {Okura}, {Tait}, {Takata}, {Tanaka}, {Tanaka},
  {Terai}, {Tomono}, {Uraguchi}, {Usuda}, {Utsumi}, {Yamada}, {Yamanoi},
  {Aihara}, {Fujimori}, {Mineo}, {Miyatake}, {Oguri}, {Uchida}, {Tanaka},
  {Yasuda}, {Takada}, {Murayama}, {Nishizawa}, {Sugiyama}, {Chiba}, {Futamase},
  {Wang}, {Chen}, {Ho}, {Liaw}, {Chiu}, {Ho}, {Lai}, {Lee}, {Jeng}, {Iwamura},
  {Armstrong}, {Bickerton}, {Bosch}, {Gunn}, {Lupton}, {Loomis}, {Price},
  {Smith}, {Strauss}, {Turner}, {Suzuki}, {Miyazaki}, {Muramatsu}, {Yamamoto},
  {Endo}, {Ezaki}, {Ito}, {Kawaguchi}, {Sofuku}, {Taniike}, {Akutsu}, {Dojo},
  {Kasumi}, {Matsuda}, {Imoto}, {Miwa}, {Suzuki}, {Takeshi}, \&
  {Yokota}}]{2018PASJ...70S...1M}
{Miyazaki}, S., {Komiyama}, Y., {Kawanomoto}, S., {et~al.} 2018, \pasj, 70, S1

\bibitem[{{Moore} {et~al.}(1999){Moore}, {Ghigna}, {Governato}, {Lake},
  {Quinn}, {Stadel}, \& {Tozzi}}]{1999ApJ...524L..19M}
{Moore}, B., {Ghigna}, S., {Governato}, F., {et~al.} 1999, \apjl, 524, L19

\bibitem[{{Moster} {et~al.}(2013){Moster}, {Naab}, \& {White}}]{moster2013}
{Moster}, B.~P., {Naab}, T., \& {White}, S. D.~M. 2013, \mnras, 428, 3121

\bibitem[{{M{\"u}ller} {et~al.}(2021){M{\"u}ller}, {Fahrion}, {Rejkuba},
  {Hilker}, {Lelli}, {Lutz}, {Pawlowski}, {Coccato}, {Anand}, \&
  {Jerjen}}]{2021A&A...645A..92M}
{M{\"u}ller}, O., {Fahrion}, K., {Rejkuba}, M., {et~al.} 2021, \aap, 645, A92

\bibitem[{{M{\"u}ller} {et~al.}(2023){M{\"u}ller}, {Heesters}, {Jerjen},
  {Anand}, \& {Revaz}}]{2023A&A...673A.160M}
{M{\"u}ller}, O., {Heesters}, N., {Jerjen}, H., {Anand}, G., \& {Revaz}, Y.
  2023, \aap, 673, A160

\bibitem[{{M{\"u}ller} \& {Jerjen}(2020)}]{2020A&A...644A..91M}
{M{\"u}ller}, O. \& {Jerjen}, H. 2020, \aap, 644, A91

\bibitem[{{M{\"u}ller} {et~al.}(2015){M{\"u}ller}, {Jerjen}, \&
  {Binggeli}}]{2015A&A...583A..79M}
{M{\"u}ller}, O., {Jerjen}, H., \& {Binggeli}, B. 2015, \aap, 583, A79

\bibitem[{{M{\"u}ller} {et~al.}(2017{\natexlab{a}}){M{\"u}ller}, {Jerjen}, \&
  {Binggeli}}]{2017A&A...597A...7M}
{M{\"u}ller}, O., {Jerjen}, H., \& {Binggeli}, B. 2017{\natexlab{a}}, \aap,
  597, A7

\bibitem[{{M{\"u}ller} {et~al.}(2018{\natexlab{a}}){M{\"u}ller}, {Jerjen}, \&
  {Binggeli}}]{2018A&A...615A.105M}
{M{\"u}ller}, O., {Jerjen}, H., \& {Binggeli}, B. 2018{\natexlab{a}}, \aap,
  615, A105

\bibitem[{{M{\"u}ller} {et~al.}(2018{\natexlab{b}}){M{\"u}ller}, {Rejkuba}, \&
  {Jerjen}}]{2018A&A...615A..96M}
{M{\"u}ller}, O., {Rejkuba}, M., \& {Jerjen}, H. 2018{\natexlab{b}}, \aap, 615,
  A96

\bibitem[{{M{\"u}ller} {et~al.}(2019){M{\"u}ller}, {Rejkuba}, {Pawlowski},
  {Ibata}, {Lelli}, {Hilker}, \& {Jerjen}}]{2019A&A...629A..18M}
{M{\"u}ller}, O., {Rejkuba}, M., {Pawlowski}, M.~S., {et~al.} 2019, \aap, 629,
  A18

\bibitem[{{M{\"u}ller} {et~al.}(2017{\natexlab{b}}){M{\"u}ller}, {Scalera},
  {Binggeli}, \& {Jerjen}}]{2017A&A...602A.119M}
{M{\"u}ller}, O., {Scalera}, R., {Binggeli}, B., \& {Jerjen}, H.
  2017{\natexlab{b}}, \aap, 602, A119

\bibitem[{{Munshi} {et~al.}(2021){Munshi}, {Brooks}, {Applebaum},
  {Christensen}, {Quinn}, \& {Sligh}}]{munshi2021}
{Munshi}, F., {Brooks}, A.~M., {Applebaum}, E., {et~al.} 2021, \apj, 923, 35

\bibitem[{{Nelson} {et~al.}(2019){Nelson}, {Springel}, {Pillepich},
  {Rodriguez-Gomez}, {Torrey}, {Genel}, {Vogelsberger}, {Pakmor}, {Marinacci},
  {Weinberger}, {Kelley}, {Lovell}, {Diemer}, \&
  {Hernquist}}]{2019ComAC...6....2N}
{Nelson}, D., {Springel}, V., {Pillepich}, A., {et~al.} 2019, Computational
  Astrophysics and Cosmology, 6, 2

\bibitem[{{Pawlowski}(2018)}]{2018MPLA...3330004P}
{Pawlowski}, M.~S. 2018, Modern Physics Letters A, 33, 1830004

\bibitem[{{Pawlowski}(2021)}]{2021NatAs...5.1185P}
{Pawlowski}, M.~S. 2021, Nature Astronomy, 5, 1185

\bibitem[{{Peng} {et~al.}(2002){Peng}, {Ho}, {Impey}, \&
  {Rix}}]{2002AJ....124..266P}
{Peng}, C.~Y., {Ho}, L.~C., {Impey}, C.~D., \& {Rix}, H.-W. 2002, \aj, 124, 266

\bibitem[{{Pritzl} {et~al.}(2003){Pritzl}, {Knezek}, {Gallagher}, {Grossi},
  {Disney}, {Minchin}, {Freeman}, {Tolstoy}, \& {Saha}}]{2003ApJ...596L..47P}
{Pritzl}, B.~J., {Knezek}, P.~M., {Gallagher}, III, J.~S., {et~al.} 2003,
  \apjl, 596, L47

\bibitem[{{Rejkuba}(2004)}]{2004A&A...413..903R}
{Rejkuba}, M. 2004, \aap, 413, 903

\bibitem[{{Revaz} \& {Jablonka}(2018)}]{2018A&A...616A..96R}
{Revaz}, Y. \& {Jablonka}, P. 2018, \aap, 616, A96

\bibitem[{{Rey} {et~al.}(2020){Rey}, {Pontzen}, {Agertz}, {Orkney}, {Read}, \&
  {Rosdahl}}]{rey2020}
{Rey}, M.~P., {Pontzen}, A., {Agertz}, O., {et~al.} 2020, \mnras, 497, 1508

\bibitem[{{Rey} {et~al.}(2019){Rey}, {Pontzen}, {Agertz}, {Orkney}, {Read},
  {Saintonge}, \& {Pedersen}}]{rey2019}
{Rey}, M.~P., {Pontzen}, A., {Agertz}, O., {et~al.} 2019, \apjl, 886, L3

\bibitem[{{Roberts} {et~al.}(2021){Roberts}, {Nierenberg}, \&
  {Peter}}]{2021MNRAS.502.1205R}
{Roberts}, D.~M., {Nierenberg}, A.~M., \& {Peter}, A. H.~G. 2021, \mnras, 502,
  1205

\bibitem[{{Saifollahi} {et~al.}(2021){Saifollahi}, {Trujillo}, {Beasley},
  {Peletier}, \& {Knapen}}]{2021MNRAS.502.5921S}
{Saifollahi}, T., {Trujillo}, I., {Beasley}, M.~A., {Peletier}, R.~F., \&
  {Knapen}, J.~H. 2021, \mnras, 502, 5921

\bibitem[{{Sales} {et~al.}(2022){Sales}, {Wetzel}, \&
  {Fattahi}}]{2022NatAs...6..897S}
{Sales}, L.~V., {Wetzel}, A., \& {Fattahi}, A. 2022, Nature Astronomy, 6, 897

\bibitem[{{Sanati} {et~al.}(2023){Sanati}, {Jeanquartier}, {Revaz}, \&
  {Jablonka}}]{sanati2023}
{Sanati}, M., {Jeanquartier}, F., {Revaz}, Y., \& {Jablonka}, P. 2023, \aap,
  669, A94

\bibitem[{{Sawala} {et~al.}(2023){Sawala}, {Cautun}, {Frenk}, {Helly},
  {Jasche}, {Jenkins}, {Johansson}, {Lavaux}, {McAlpine}, \&
  {Schaller}}]{2023NatAs...7..481S}
{Sawala}, T., {Cautun}, M., {Frenk}, C., {et~al.} 2023, Nature Astronomy, 7,
  481

\bibitem[{{Sawala} {et~al.}(2016{\natexlab{a}}){Sawala}, {Frenk}, {Fattahi},
  {Navarro}, {Bower}, {Crain}, {Dalla Vecchia}, {Furlong}, {Helly}, {Jenkins},
  {Oman}, {Schaller}, {Schaye}, {Theuns}, {Trayford}, \&
  {White}}]{2016MNRAS.457.1931S}
{Sawala}, T., {Frenk}, C.~S., {Fattahi}, A., {et~al.} 2016{\natexlab{a}},
  \mnras, 457, 1931

\bibitem[{{Sawala} {et~al.}(2016{\natexlab{b}}){Sawala}, {Frenk}, {Fattahi},
  {Navarro}, {Bower}, {Crain}, {Dalla Vecchia}, {Furlong}, {Helly}, {Jenkins},
  {Oman}, {Schaller}, {Schaye}, {Theuns}, {Trayford}, \& {White}}]{sawala2016}
{Sawala}, T., {Frenk}, C.~S., {Fattahi}, A., {et~al.} 2016{\natexlab{b}},
  \mnras, 457, 1931

\bibitem[{{Schneider}(2015)}]{2015MNRAS.451.3117S}
{Schneider}, A. 2015, \mnras, 451, 3117

\bibitem[{{Simon} \& {Geha}(2007)}]{2007ApJ...670..313S}
{Simon}, J.~D. \& {Geha}, M. 2007, \apj, 670, 313

\bibitem[{{Simpson} {et~al.}(2018){Simpson}, {Grand}, {G{\'o}mez}, {Marinacci},
  {Pakmor}, {Springel}, {Campbell}, \& {Frenk}}]{2018MNRAS.478..548S}
{Simpson}, C.~M., {Grand}, R.~J.~J., {G{\'o}mez}, F.~A., {et~al.} 2018, \mnras,
  478, 548

\bibitem[{{Smercina} {et~al.}(2018){Smercina}, {Bell}, {Price}, {D'Souza},
  {Slater}, {Bailin}, {Monachesi}, \& {Nidever}}]{2018ApJ...863..152S}
{Smercina}, A., {Bell}, E.~F., {Price}, P.~A., {et~al.} 2018, \apj, 863, 152

\bibitem[{{Springel} {et~al.}(2018){Springel}, {Pakmor}, {Pillepich},
  {Weinberger}, {Nelson}, {Hernquist}, {Vogelsberger}, {Genel}, {Torrey},
  {Marinacci}, \& {Naiman}}]{2018MNRAS.475..676S}
{Springel}, V., {Pakmor}, R., {Pillepich}, A., {et~al.} 2018, \mnras, 475, 676

\bibitem[{{Taylor} {et~al.}(2018){Taylor}, {Eigenthaler}, {Puzia}, {Mu{\~n}oz},
  {Ribbeck}, {Zhang}, {Ordenes-Brice{\~n}o}, \& {Bovill}}]{2018ApJ...867L..15T}
{Taylor}, M.~A., {Eigenthaler}, P., {Puzia}, T.~H., {et~al.} 2018, \apjl, 867,
  L15

\bibitem[{{Taylor} {et~al.}(2017){Taylor}, {Puzia}, {Mu{\~n}oz}, {Mieske},
  {Lan{\c c}on}, {Zhang}, {Eigenthaler}, \& {Bovill}}]{2017MNRAS.469.3444T}
{Taylor}, M.~A., {Puzia}, T.~H., {Mu{\~n}oz}, R.~P., {et~al.} 2017, \mnras,
  469, 3444

\bibitem[{{Tollerud} {et~al.}(2008){Tollerud}, {Bullock}, {Strigari}, \&
  {Willman}}]{2008ApJ...688..277T}
{Tollerud}, E.~J., {Bullock}, J.~S., {Strigari}, L.~E., \& {Willman}, B. 2008,
  \apj, 688, 277

\bibitem[{{Tonry} \& {Schneider}(1988)}]{1988AJ.....96..807T}
{Tonry}, J. \& {Schneider}, D.~P. 1988, \aj, 96, 807

\bibitem[{{van Dokkum} {et~al.}(2018){van Dokkum}, {Danieli}, {Cohen},
  {Merritt}, {Romanowsky}, {Abraham}, {Brodie}, {Conroy}, {Lokhorst}, {Mowla},
  {O'Sullivan}, \& {Zhang}}]{2018Natur.555..629V}
{van Dokkum}, P., {Danieli}, S., {Cohen}, Y., {et~al.} 2018, \nat, 555, 629

\bibitem[{{Wang} {et~al.}(2015){Wang}, {Dutton}, {Stinson}, {Macci{\`o}},
  {Penzo}, {Kang}, {Keller}, \& {Wadsley}}]{wang2015}
{Wang}, L., {Dutton}, A.~A., {Stinson}, G.~S., {et~al.} 2015, \mnras, 454, 83

\bibitem[{{Weinberg} {et~al.}(2015){Weinberg}, {Bullock}, {Governato}, {Kuzio
  de Naray}, \& {Peter}}]{2015PNAS..11212249W}
{Weinberg}, D.~H., {Bullock}, J.~S., {Governato}, F., {Kuzio de Naray}, R., \&
  {Peter}, A. H.~G. 2015, Proceedings of the National Academy of Science, 112,
  12249

\bibitem[{{Wetzel} {et~al.}(2016{\natexlab{a}}){Wetzel}, {Hopkins}, {Kim},
  {Faucher-Gigu{\`e}re}, {Kere{\v s}}, \& {Quataert}}]{2016ApJ...827L..23W}
{Wetzel}, A.~R., {Hopkins}, P.~F., {Kim}, J.-h., {et~al.} 2016{\natexlab{a}},
  \apjl, 827, L23

\bibitem[{{Wetzel} {et~al.}(2016{\natexlab{b}}){Wetzel}, {Hopkins}, {Kim},
  {Faucher-Gigu{\`e}re}, {Kere{\v{s}}}, \& {Quataert}}]{wetzel2016}
{Wetzel}, A.~R., {Hopkins}, P.~F., {Kim}, J.-h., {et~al.} 2016{\natexlab{b}},
  \apjl, 827, L23

\bibitem[{{Wheeler} {et~al.}(2019){Wheeler}, {Hopkins}, {Pace},
  {Garrison-Kimmel}, {Boylan-Kolchin}, {Wetzel}, {Bullock}, {Kere{\v{s}}},
  {Faucher-Gigu{\`e}re}, \& {Quataert}}]{wheeler2019}
{Wheeler}, C., {Hopkins}, P.~F., {Pace}, A.~B., {et~al.} 2019, \mnras, 490,
  4447

\bibitem[{{Wheeler} {et~al.}(2015){Wheeler}, {O{\~n}orbe}, {Bullock},
  {Boylan-Kolchin}, {Elbert}, {Garrison-Kimmel}, {Hopkins}, \&
  {Kere{\v{s}}}}]{wheeler2015}
{Wheeler}, C., {O{\~n}orbe}, J., {Bullock}, J.~S., {et~al.} 2015, \mnras, 453,
  1305

\end{thebibliography}

\appendix

\section{Dwarf galaxy candidate dw1336-32}
\label{app:dw1336-32}
\begin{figure}
  \begin{center}
  \includegraphics[width=\linewidth]{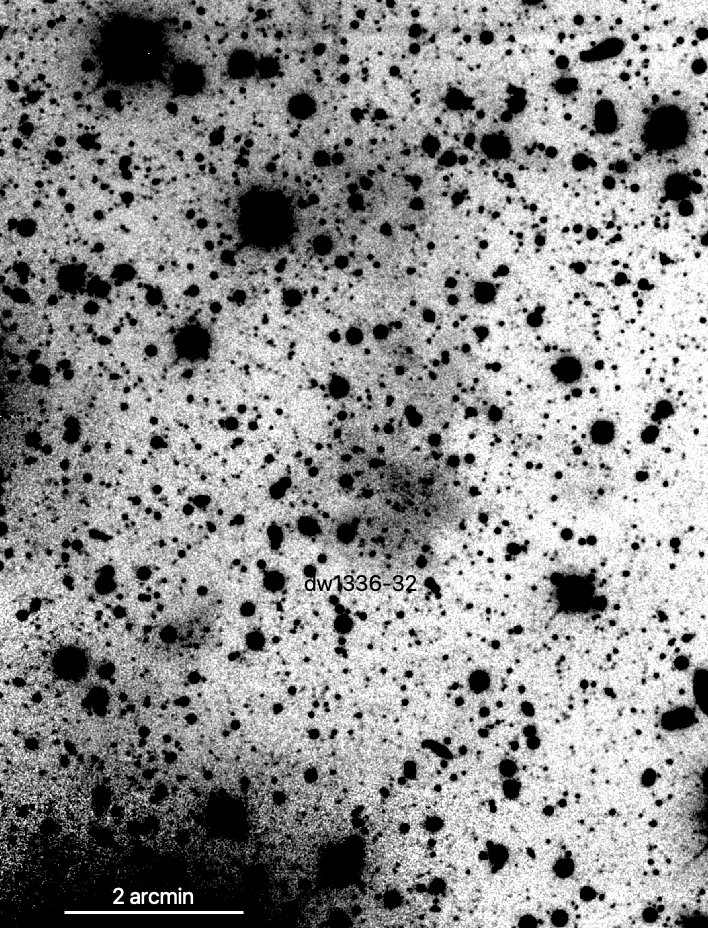}
  \caption{Deep single band exposure of the dwarf galaxy candidate dw1336-32 in the $g$ band, smoothed with a Gaussian kernel.}
  \label{cirrus}
  \end{center}
\end{figure}

{
In Fig\,\ref{cirrus} we show a 6,000\,s exposure in the $g$ band with DECam (from the program id 2016A-0384) of the dwarf galaxy candidate dw1336-32 from \citet{2015A&A...583A..79M}. It has a patchy morphology and is embedded in a cirrus-like structure, making it a likely false detection. No surface brightness fluctuation is apparent. We reject it as a dwarf galaxy.
 }

\end{document}